\documentclass[12pt,twoside]{article}
\usepackage{psfig}
\usepackage{amssymb}
\usepackage{graphicx}
\begin{document}
\begin{center}
%\documentstyle[12pt,twoside]{article}
%\evensidemargin=0mm
%\oddsidemargin=0mm
%\topmargin=-7mm
%\textwidth=160mm
%\textheight=225mm
%\begin{document}
%\begin{center}
{\Large\bf
DARK ENERGY, DARK MATTER AND\\[5PT]
THE CHAPLYGIN GAS\\[5PT]}
\medskip
{\bf
R.Colistete Jr.\footnote{e-mail: coliste@ccr.jussieu.fr},J. C. Fabris\footnote{e-mail: fabris@cce.ufes.br},
S.V.B. Gon\c{c}alves\footnote{e-mail: sergio@cce.ufes.br} and P.E. de Souza\footnote{e-mail: patricia.ilus@bol.com.br}}\\  \medskip
Departamento de F\'{\i}sica, Universidade Federal do Esp\'{\i}rito Santo,
CEP29060-900, Vit\'oria, Esp\'{\i}rito Santo, Brazil \medskip
\end{center}
\begin{abstract}
The possibility that the dark energy may be described by the Chaplygin gas is discussed.
Some observational constraints are established. These observational
constraints indicate that a unified model for dark energy and dark matter through the
employement of the Chaplygin gas is favored.
\vspace{0.7cm}
\newline
PACS number(s): 98.80.Bp, 98.65.Dx
\end{abstract}

The possible existence of dark matter and dark energy in the Universe constitutes one
of the most important problems in physics today. One calls dark matter the mysterious
component of galaxies and clusters of galaxies that manifestates itself only through
gravitational effects, affecting only the dynamics of those object. Dark matter appears
in collapsed objects. The existence of dark energy, on the other hand, has been speculated
due to the observation of the Universe as a whole. In particular, the position of the
first accoustic peak in the spectrum of the anisotropy of the cosmic microwave background
radiation\cite{charles} and the data from supernova type Ia\cite{riess,perlmutter} suggest that the dynamics of the Universe
is such that the most important matter component is non-clustered, exhibiting negative
pressure. The current observations favor $\Omega_{dm} \sim 0.3$ and
$\Omega_{de} \sim 0.7$
for the proportion of dark matter and dark energy.
\par
The most natural candidate to represent dark energy is a cosmological constant\cite{carroll}. However,
it is necessary a fine tunning of $120$ orders of magnitude in order to obtain agreement
with observations. Another popular candidate today is quintessence, a self-interacting
scalar field\cite{steinhardt,sahni,brax}. But the quintessence program suffers from fine tuning of
microphysical paremeters.
In this work, we discuss the possibility that the dark energy may be represented by
the Chaplygin gas\cite{pasquier,patricia,patricia2,bilic,bento}, which is characterized by the equation of state
\begin{equation}
\label{eos}
p = - \frac{A}{\rho} \quad .
\end{equation}
\par
The Chaplygin gas has an interesting motivation connected with string theory. In fact,
if we consider a $d$-brane configuration in the $d+2$ Nambu-Goto action, the  employement
of the light-cone parametrization leads to the action of a newtonian fluid with the
equation of state (\ref{eos}), whose symmetries are the same as the Poincar\'e group.
Hence, the relativistic character of the action is somehow hidden in the
equation of state (\ref{eos}). For a review of the properties of the Chaplygin gas
see reference \cite{jackiw}.
\par
Using the relativistic equation of conservation for a fluid in a FRW background,
we obtain that the density of the Chaplygin gas depends on the scale factor as
\begin{equation}
\rho = \sqrt{A + \frac{B}{a^6}} \quad .
\end{equation}
Hence, initially the Chaplygin gas behaves as a pressurelless fluid and assumes
asymptotically a behaviour typical of a cosmological constant. In this sense,
the Chaplygin gas may be an alternative for the description of dark energy. Remark that,
even if the pressure associated with the Chaplygin gas is negative, the sound velocity
is real. Hence, no instability problem occurs as it happens with other fluids with
negative pressure\cite{jerome}.
\par
In reference \cite{patricia} the density perturbations in a Universe dominated by the Chaplygin
gas was investigate. In that work, the fact that the Chaplygin gas may be represented
by a newtonian fluid was explored. The density contrast behaves as
\begin{equation}
\label{fs}
\delta = t^{-1/6}\biggr[C_1J_\nu(\Sigma^2nt^{7/3})
+ C_2J_{-\nu}(\Sigma^2nt^{7/3})\biggl]\quad,
\end{equation}
where $n$ is the wavenumber of the perturbations, $\nu = 5/14$ and $\Sigma^2 = \frac{54}{49}\frac{A\pi G}{a_0^2}$, $a_0$ is a constant and
$C_1$ and $C_2$ are constants that depend on $n$. A simple analysis of
this solution indicates that initially the perturbations behave as in the case of
the pressurelless fluid, presenting asymptotically oscillatory behaviour with
decreasing amplitude, approaching a zero value typical of a cosmological constant.
Even if this result was obtained in a newtonian context, a numerical analysis of the
corresponding relativistic equations exhibit the same behaviour.
\par
In reference \cite{patricia2}, the mass power spectrum of the perturbations in a Universe dominated
by the Chaplygin gas was computed. The matter content was assumed to be composed of
radiation, dark matter and dark energy, the latter represented by the Chaplygin gas.
Due to the complexity of perturbed equations, a numerical analysis was performed.
The initial spectrum, at the moment of decoupling of radiation and matter, was supposed
to be the Harrisson-Zeldovich scale invariant spectrum. Taken at constant time, this
implies to impose an initial spectral index $n_s = 5$. The system evolves and the power spectrum is computed for the present time.
The final results were obtained for a Universe dominated uniquely by a pressurelless
fluid ({\it baryonic model}), for a cosmological constant model and for a mixed of dark matter and
Chaplygin gas, with different values for the sound velocity of the Chaplygin gas,
defined as $v^2_s = \bar A = A/\rho_{c0}^2$, where $\rho_{c0}$ is the Chaplygin gas
density today. When $\bar A = 1$ the Chaplygin gas becomes identical to a cosmological constant. It was verified that the Chaplygin gas models interpolate the baryonic model
and the cosmological constant model as $\bar A$ varies from zero to unity. The results
for the power spectrum
$P(n) = n^{3/2}\delta_n$ in terms of $n/n_0$, where $n_0$ is reference scale corresponding
to $100\,Mpc$, are displayed in figure $1$ for the baryonic model, cosmological constant model and
for the Chaplygin gas model with different values for $\bar A$. The spectral indice
for perturbations of some hundreds of megaparsecs up to the Hubble radius
is $n_s \sim 4.7$ for the baryonic model, $n_s \sim 4.2$ for the cosmological constant
model, $n_s \sim 4.5$ for the Chaplygin gas with $\bar A = 0.5$.
\par
Another important test to verify if the Chaplygin gas model may represent dark energy
is the comparison with the supernova type Ia data. In order to do so, we evaluate
the luminosity distance\cite{coles} in the Chaplygin gas cosmological model. In such a model,
the luminosity distance, for a flat Universe, reads
\begin{equation}
D_L = (1+z)\int_0^z\frac{dz'}{\sqrt{\Omega_{m0}(1 + z')^3 +
\Omega_{c0}\sqrt{\bar A + (1 - \bar A)(1 + z')^6}}} \quad ,
\end{equation}
$z$ being the redshift.
From this expression the following relation
between the apparent magnitude $m$ and the absolute magnitude $M$ is obtained:
\begin{equation}
m - M = 5\log\biggr\{(1 + z)\int_0^z\frac{dz'}{\sqrt{\Omega_{m0}(1 + z')^3 +
\Omega_{c0}\sqrt{\bar A + (1 - \bar A)(1 + z')^6}}}\biggl\} \quad .
\end{equation}
\par
In order to compare with the supernova data, we compute the quantity
\begin{equation}
\mu_0 = 5\log\biggr(\frac{D_L}{Mpc}\biggl) + 25 \quad.
\end{equation}
\par
The quality of the fitting is characterized by the parameter
\begin{equation}
\chi^2 = \sum_i\frac{[\mu_{0,i}^o - \mu_{0,i}^t]^2}{\sigma_{\mu_0,i}^2 + \sigma_{mz,i}^2}
\quad .
\end{equation}
In this expression, $\mu_{0,i}^o$ is the measured value, $\mu_{0,i}^t$ is the value
calculated through the model described above, $\sigma_{\mu_0,i}^2$ is the measuremente
error, $\sigma_{mz,i}^2$ is the dispersion in the distance modulus due to the dispersion
in galaxy redshift due to perculiar velocities. This quantity we will taken as
\begin{equation}
\sigma_{mz} = \frac{\partial \log D_L}{\partial z}\sigma_z \quad ,
\end{equation}
where, following \cite{perlmutter}, $\sigma_z = 200\,km/s$.
We evaluate, in fact, $\chi^2_\nu$, the estimated errors for degree of freedom.
\par
The best fitting was obtained for the case where $\Omega_{m0} = 0$, $\Omega_{c0} = 1$,
$\bar A \sim 0.847$, $H_0 \sim 62.1$, with $T_0 \sim 15.1\,Gy$ for the age of Universe.
In this case, $\chi^2_\nu \sim 0.760$.
As a comparison, if we substitute the Chaplygin gas by a cosmological constant
($\bar A = 1$), the best fitting is given by $\Omega_{m0} \sim 0.244$, $\Omega_\Lambda \sim
0.756$, $H_0 \sim 61.8$, and the age of the Universe is $T_0 \sim 16.5\,Gy$. In this
model $\chi^2_\nu \sim 0.774$. In figure $2$ we display this fitting.
\par
These results deserve some comments. First, if we consider a cosmological constant
model, the best fitting is in quite good agreement with other observational data,
mainly those from $CMB$ and the dynamics of clusters of galaxies. However, the
Chaplygin gas model indicates no need of dark matter, and the fitting quality is slightly
better than the cosmological model. This is interesting since there are claims that
the Chaplygin gas may unify dark matter with dark energy: at small scales, the Chaplygin
gas may cluster, playing the role of dark matter, remaining a smooth component at large
scale \cite{pasquier,bilic,bento}.
\par
The results presented above indicate that the Chaplygin gas may be taken seriously as
a candidate to describe dark energy and dark matter. However, more analysis are needed
in order to verify to which extent it can be a better alternative with respect to
the cosmological constant model combined with dark energy. A crucial test is the analysis
of the spectrum of anisotropy of the cosmic microwave background radiation.

\newpage

\begin{figure}[tb]
\begin{center}
\includegraphics[scale=0.8]{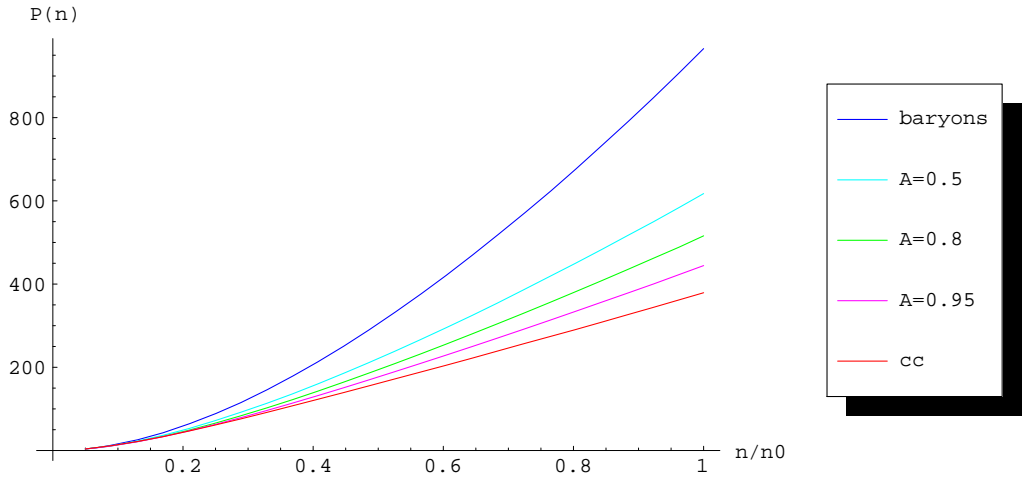}
\end{center}
\caption{{\protect\footnotesize
The power spectrum $P(n) = n^{3/2}\delta_n$ as function of $n/n_0$.
}}
\label{fig1}
\end{figure}

\begin{figure}[tb]
\begin{center}
\includegraphics[scale=0.8]{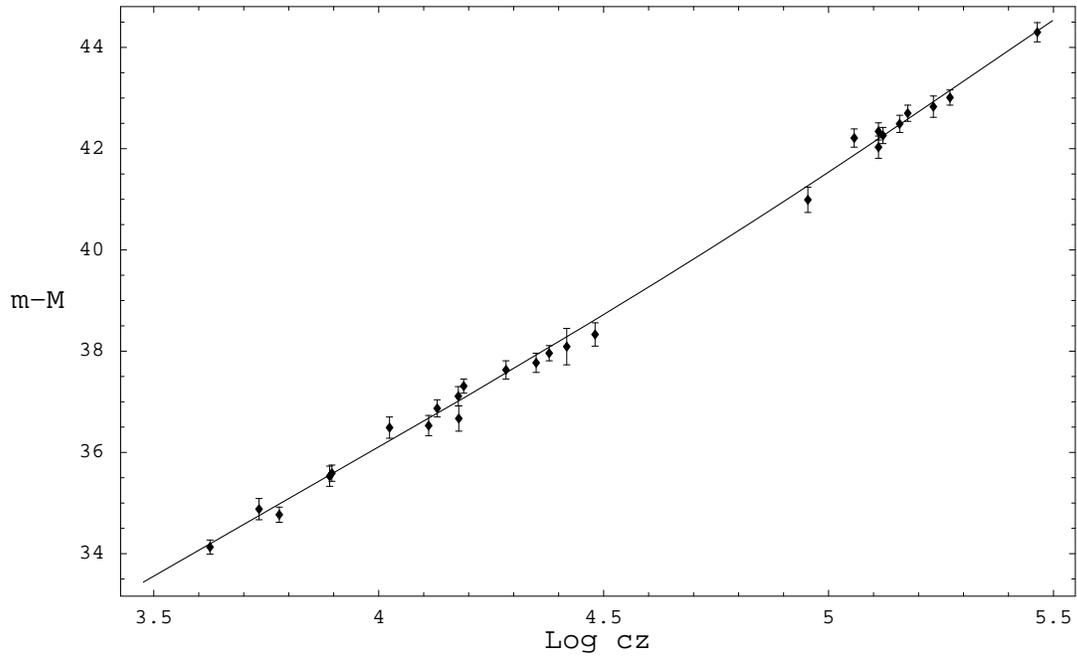}
\end{center}
\caption{{\protect\footnotesize
The the best fitting model, where $\Omega_{m0} = 0$, $\Omega_{c0} = 1$, $\bar A \sim
0.847$ and $H_0 \sim 62.1$.
}}
\label{fig2}
\end{figure}


\begin{thebibliography}{100}
\bibitem{charles} C.H. Lineweaver, {\it Cosmological parameters}, astro-ph/0112381;
\bibitem{riess} A.G. Riess et al, Astron.J. {bf 116}, 1009(1998);
\bibitem{perlmutter} S. Permultter et al, Astrophys. J. {\bf 517}, 565(1998);
\bibitem{carroll} S.M. Carroll, Liv. Rev. Rel. {\bf 4}, 1(2001);
\bibitem{steinhardt} R.R. Caldwell, R. Dave and P. Steinhardt, Phys. Rev. Lett. {\bf 80}, 1582(1998);
\bibitem{sahni} V. Sahni, {\it The cosmological constant problem and quintessence}, astro-ph/0202076;
\bibitem{brax} Ph. Brax and J. Martin, Phys. Lett. {\bf B468}, 40(1999);
\bibitem{pasquier} A. Kamenshchik, U. Moschella and V. Pasquier, Phys. Lett. {\bf B511}, 265(2001);
\bibitem{patricia} J.C. Fabris, S.V.B. Gon\c{c}alves and P.E. de Souza, Gen. Rel. Grav.
{\bf 34}, 53(2002);
\bibitem{patricia2} J.C. Fabris, S.V.B. Gon\c{c}alves e P.E. de Souza, {\it Mass power spectrum
in a Universe dominated by the Chaplygin gas}, astro-ph/0203441;
\bibitem{bilic} N. Bilic, G.B. Tupper and R.D. Viollier, Phys. Lett. {\bf B535}, 17(2002);
\bibitem{bento} M.C. Bento, O. Bertolami and A.A. Sen, {\it Generalized Chaplygin gas, accelerated expansion
and dark energy-matter unification}, gr-qc/0202064;
\bibitem{jackiw} R. Jackiw, {\it A particle field theorist's
lectures on supersymmetric, non-abelian fluid mechanics and d-branes}, physics/0010042;
\bibitem{jerome} J.C. Fabris and J. Martin, Phys. Rev. {\bf D55}, 5205(1997);
\bibitem{coles} P. Coles and F. Lucchin, {\bf Cosmology}, Wiley, New York(1995).
\end{thebibliography}
\end{document}